--
\documentclass[aps,prd,twocolumn,groupedaddress]{revtex4-1}

\usepackage{graphicx}

\begin{document}


\title{Planck scale effect 
on the thermodynamics of photon gas}


\author{Mir Mehedi Faruk}
\email[]{muturza3.1416@gmail.com, mir.faruk15@imperial.ac.uk, mir.mehedi.faruk@cern.ch} 
\affiliation{Theoretical Physics, Blackett Laboratory, Imperial College, London SW7 2AZ, United Kingdom}

\author{Md. Muktadir Rahman}
\affiliation{Department of Theoretical Physics, University of Dhaka, Dhaka-1000}


\date{\today}

\begin{abstract}
A particular framework for quantum gravity is the doubly special relativity (DSR) formalism that
introduces a new observer independent scale (the Planck scale).
We resort to the methods of statistical mechanics in this framework to determine how the 
deformed dispersion relation affects the thermodynamics of a photon gas. 
The ensuing modifications to the density of states,
partition function, pressure, internal energy, entropy, free energy and specific heat are
calculated.
These results are compared with the 
outcome obtained in the Lorentz violating model of Camacho and Marcias (Gen. Relativ. Gravit. 39: 1175-1183, 2007).
The two types of models predict different results due to different spacetime structure near the Planck scale.
The resulting modifications can be interpreted as a consequence of the deformed Lorentz symmetry present in the particular model we have considered. In the low energy limit, 
our calculation coincides with usual results of photon
thermodynamics in special relativity (SR) theory,
in contrast to the study presented in  (Phys. Rev. D81, 085039 (2010)).
\end{abstract}

\pacs{02.40.Gh, 03.30.+p, 11.30.Cp}

\maketitle

\section{Introduction}
A simple paradox confronts us as we seek the quantum theory of gravity. 
The combination of gravity ($G$), the quantum  ($h$), and relativity ($c$)
gives rise to the Planck length, $l_P$, or its inverse, the Planck energy $E_P$\cite{mag1,mag2}. 
These scales mark thresholds beyond which the old description of spacetime breaks
down and qualitatively new phenomena are expected to appear. But this proposition obviously opposes
the principle of  special relativity (SR) theory, where,
the length 
of an object varies for separate observers. 
Thus,  an extension of SR theory is needed where along with the velocity of light, 
another observer-independent quantity, a fundamental length scale, exists. As a 
result, we must  modify SR theory near the high energy (Planck energy). One such modified theory is doubly special relativity (DSR)\cite{Am,Am1}, which
has drawn a lot of interest as a
possible framework of quantum gravity\cite{Am2,g}. There are mainly two basic principles 
on which this theory rests: (i) the appearance of a second observer independent scale\cite{mag1},
which can be the Planck length;  
and (ii) a naturally emerging Non-Commutative (NC) spacetime\cite{k,pal}. 
All the models of quantum gravity predict qualitatively different spacetime beyond 
a certain energy (length) scale, generally considered to be the Planck energy (length). 
Also, it is now well established\cite{m,emt,pp} that a consistent marriage of ideas of quantum mechanics and gravitation 
need a noncommutative description of spacetime to avoid the paradoxical
situation of creation of a black hole for an event that 
is sufficiently localized in spacetime. So, DSR theory fits the criterion 
for being a  quantum gravity framework in this respect\cite{mag2,k}.
Recently, DSR has also developed for curved space\cite{curved}.\\\\
Amelino-Camelia\cite{Am1} 
first proposed a possible solution to this problem.
Another model, which is much simpler, was given by  Magueijo and Smolin\cite{mag1}, and is
referred to as MS model in this paper.
The stated paradox can be solved if the Lorentz transformations can be modified
so as to preserve a single energy or momentum scale. It has been reported that it is possible\cite{pal}
to build models  keeping the principle of relativity for inertial frames intact, and to simply modify 
the laws by which energy and momenta measured by different inertial observers are related to each other.
By adding nonlinear terms to the action of the Lorentz transformations on momentum space, one can maintain 
the relativity of inertial frames. And, then all observers will agree that there is an invariant energy or
momentum above which the picture of spacetime as a smooth manifold breaks down. 
Because there are two constants which are preserved, this theory is named “doubly special relativity” 
(DSR)\cite{J. Kowalski-Glikman}. 
So, added nonlinear
terms to the action of Lorentz transformation  makes it possible to maintain the relativity of inertial
frames and solve the paradox at the same time,
but the quadratic invariant of SR  is now replaced by a nonlinear
invariant, which in turn leads to a modified dispersion relation.
In MS model,
 the Lorentz  algebra is still 
not deformed and there are no deformations in 
the brackets of rotations and momenta\cite{mag1,pal}. Briefly, 
DSR theory possesses the following 
simple but strong features:
(i) First of all, in DSR the relativity of inertial frames, as proposed by Galileo, Newton and Einstein, is well preserved. 
(ii) Secondly, there is an invariant energy scale $\kappa$, which is of the order of the Planck scale. (iii) Thirdly, in  DSR theories, the notion of absolute locality should be replaced by 
relative locality as due to the presence of an energy-dependent metric, different observers live in different spacetime\cite{kkk}.

The well-known dispersion relation
(or mass-shell condition) for a particle\cite{mag1,mag2}
in SR theory is, $(\hbar=c=1)$
\begin{eqnarray}
 E^2=p^2+m^2,
\end{eqnarray}
which has to be modified in the MS model as\cite{mag1}-
\begin{eqnarray}
 E^2=p^2+m^2(1-\frac{E}{\kappa})^2.
\end{eqnarray}
Here $E$ and $p$ are the energy and the magnitude of the three-momentum of the particle respectively, while $m$
is the mass of the particle.
We will refer to this model as the Magueijo-Smolin (MS) model in this manuscript. 
A lot of of studies have been carried out this model including analogue gravity\cite{lll}, noncommutative geometry\cite{nc},
Bose-Einstein condensate\cite{tc},
relativistic thermodynamics\cite{emt,malu, malu2},   cosmology\cite{rainbow} as well as
DSR formalism from conformal group\cite{dragon}. Recently, a lot of theoretical study have been done on 
thermodynamics of relativistic quantum gas\cite{qg} as it plays a crucial role in cosmology\cite{L,L1},
as well as in 
condensed matter\cite{mmf}.  A study on the thermodynamics of a photon gas with an invariant energy scale using the MS model has 
already been reported by Das and Roychowdhury\cite{photon}. 
In their paper, they have constructed the formalism to do such calculations within MS model. 
But there is a severe error in their 
calculation, as they have used Maxwell-Boltzmann statistics while calculating the partition
function of photons, but  it is  well known that, photons  are  integer spin quantum particles\cite{peskin}. So,
they must obey Bose Einstein distribution.
As a consequence, one must use Bose-Einstein statistics to calculate the thermodynamics of 
photon. Due to this serious error, the results obtained by Das and Roychowdhury\cite{photon}
do not coincide with known results of the thermodynamic
quantities of photon gas in SR theory\cite{pathria,ca}. For instance, their obtained internal energy of photon gas depends upon temperature, $T$
linearly, but it is well known that the internal energy $E\propto T^4$ (Stefan-Boltzman law)\cite{pathria}. 
An important
point to note is that, 
thermodynamics for photon gas with a different
dispersion relation has been studied  by Camacho and Marcias\cite{cam},
where Bose-Einstein statistics has been used as expected. Besides, the thermodynamics 
of massive bosons and fermions 
with another different dispersion relation 
has  also been investigated\cite{de}.
But both of these two
modified dispersion relations appear from a
phenomenological point of view whereas the dispersion relation (2)
has a more theoretical motivation. 

Due to its very fundamental role in theoretical physics, Lorentz symmetry has been subjected to some
of the 
highest precision tests\cite{exp,exp1,exp2}.
It has been advocated by a number
of physicists\cite{Am1,Am2} that,
Lorentz invariance (both global and
local) is only an approximate symmetry, which is broken
at the Planck scale.
Camacho and Marcias\cite{cam} examined the consequence of Lorentz violation 
in their Lorentz violating model
in a unique but different approach
where they introduced a deformed dispersion relation as a fundamental fact for the dynamics of photons
and analyzed the effects of this upon
 the thermodynamics of photon gas.
They
showed that
the breakdown of Lorentz symmetry entails an increase in the number of microstates, and  as a consequence a growth of
the entropy and other thermodynamic quantities, with respect to the case of SR theory is observed.
So, it will be really intriguing to check the status of 
the thermodynamic quantities of photons in the MS model, where the 
relativity of inertial frames, as proposed by Galileo, Newton and Einstein, is well preserved but at the
same time solves the paradox related to the 
appearance of a second observer independent scale\cite{mag1}.
The current paper is organised as follows.
In section 2, we review  shortly 
the  nonlinear realization of Lorentz group, which 
gives rise to the modified dispersion relation of Eq. (2).
In the next section, we discuss about the density of states and calculate the 
partition function.
In the section 4, 
we go on to study the
thermodynamic properties of photon gas using the derived
partition function. We do the whole calculation in arbitrary dimensions
but specially scrutinize the thermodynamic properties for three dimensional space.
The different relations between the thermodynamic quantities of photons in SR theory, such
as the pressure-energy density relation and the entropy-specific heat relation do not remain valid in the Lorentz-violating
model of Camacho and Marcias\cite{cam}. We carefully check if these identities are still valid in the MS model, where Lorentz symmetry is still
preserved.

\section{Non linear realization of Lorentz algebra and MODIFIED DISPERSION RELATION}
In this section, while 
working in MS model\cite{mag1} we briefly review\cite{pal} the 
non linear realization of Lorentz algebra in $(d+1)$
dimensional spacetime.
Interested reader can go through \cite{pal}
for more details. 
Starting  from the familiar (linear) SR Lorentz transformation, the $L_{SR}$ coordinate space variable,
\begin{eqnarray}
&&X'^0=L_{SR}(X^0)=\gamma(X^0-\nu X^1),\nonumber\\
&&X'^1=L_{SR}(X^1)=\gamma(X^1-\nu X^0),\nonumber\\
&&X'^2=L_{SR}(X^2)=X^2,\nonumber\\
&&X'^3=L_{SR}(X^3)=X^3,\\
&&.....................\nonumber\\
&&.....................\nonumber\\
&&X'^d=L_{SR}(X^d)=X^d,\nonumber
\end{eqnarray}
where, $\gamma=(1-\nu^2)^{-1/2}$ and the boost is along the  $X^1$ direction with velocity $\nu^i=(\nu,0,0,..,0)$.
Continuing in the same way for the momentum space variable we get,
\begin{eqnarray}
 &&P'^0=L_{SR}(P^0)=\gamma(P^0-\nu P^1),\nonumber\\
&&P'^1=L_{SR}(P^1)=\gamma(P^1-\nu P^0),\nonumber\\
&&P'^2=L_{SR}(P^2)=P^2,\nonumber\\
&&P'^3=L_{SR}(P^3)=P^3,\\
&&.....................\nonumber\\
&&.....................\nonumber\\
&&P'^d=L_{SR}(P^d)=P^d.\nonumber
\end{eqnarray}
In eq. (2) and (3),  
($X^\mu,P^\mu$) are the  phase space variables who obey 
normal Poisson
Bracket algebra,
commuting or more precisely, canonical degrees of freedom.
Let us now declare 
$\kappa$-Minkowski phase space elements $(x^\mu,p^\mu)$, where $x^\mu$ and $p^\mu$ are the position and momentum space coordinates 
respectively, who satisfy non commutative the $\kappa$-Minkowski phase space algebra and
DSR-Lorentz transformations\cite{pal}.
Now defining an invertible map $F$ such that\cite{pal,emt},
\begin{eqnarray}
&& F(X^\mu)\longrightarrow x^\mu,\nonumber\\
&&F^{-1}(x^\mu)\longrightarrow X^\mu,
 \end{eqnarray}
which in explicit form reads:
\begin{eqnarray}
 F(X^\mu)=x^\mu(1-\frac{p^0}{\kappa}),\\
F^{-1}(x^\mu)=X^\mu(1+\frac{p^0}{\kappa}),\\
 F(P^\mu)=\frac{p^\mu}{(1-\frac{P^0}{\kappa})},\\
 F^{-1}(p^\mu)=\frac{P^\mu}{(1+\frac{P^0}{\kappa})}.
 \end{eqnarray}
Now the DSR-Lorentz transformation $L_{DSR}$ is formally expressed as,
\begin{eqnarray}
 x'^\mu=L_{DSR}(x^\mu)=F\circ L_{SR}\circ F^{-1}(x^\mu),\\
  p'^\mu=L_{DSR}(p^\mu)=F\circ L_{SR}\circ F^{-1}(p^\mu).
\end{eqnarray}
In the case of $x^0$,
\begin{eqnarray}
 x'^0 &=&L_{DSR}(x^0)=F\circ L_{SR}\circ F^{-1}(x^0)\nonumber\\
 &=& F\circ L_{SR}(X^0(1+\frac{P^0}{\kappa}))\nonumber\\
 &=& F(\gamma(X^0-\nu X^1)(1+\frac{\gamma}{\kappa}(P^0-\nu P^1) )\nonumber \\
 &=& \gamma\alpha (x^0-\nu x^1),
\end{eqnarray}
where, 
\begin{eqnarray}
 \alpha=1 +\kappa^{-1}( (\gamma-1)P^0-\gamma \nu P^1 ),
\end{eqnarray}
In the same way we find out,
\begin{eqnarray}
 x'^1=\gamma\alpha(x^1-\nu x^0),\\
p'^0 =\frac{\gamma}{\alpha}(p^0-\nu p^1),\\
 p'^1=\frac{\gamma}{\alpha}(p^1-\nu p^0).
 \end{eqnarray}
And  transverse component of $x^\mu$
and $p^\mu$
transforms as,
\begin{eqnarray}
 x'^i=\alpha x^i\\
 p'^i =\frac{p^i}{\alpha}
\end{eqnarray}
where, $i=2,3, .., d$.
It is very interesting to see how 
in the present formulation \cite{pal},
noncommutative effects enter through these generalized
(nonlinear) transformation rules.
Most importantly, the transverse components
 also transform due to nonlinear realization of the Lorentz group, unlike the usual SR transformation (eq (3)).
As expected, in the limit $\kappa\rightarrow\infty$, the generalized  transformation rule coincides with 
SR transformation.
Therefore, the phase space quantity invariant under DSR-Lorentz transformation is-
$\eta_{\mu\nu} p^{\mu}p^\nu(1-\frac{p^0}{\kappa})^{-2}$, where $\eta_{\mu\nu}=diag(-1,1,1,...,1)$,
writing this as,
\begin{eqnarray}
 m^2=\eta_{\mu\nu} p^{\mu}p^\nu(1-\frac{p^0}{\kappa})^{-2}
\end{eqnarray}
This yields the well-known dispersion relation due to Magueijo and Smolin in Eq (2).
It is shown in \cite{velo} that a modified
dispersion relation does not necessarily imply a varying
(energy dependent) velocity of light. But there
are models\cite{cam,de} that admit a varying speed of light.
Although,
in the case of the MS model, for photons ($m = 0$) the
dispersion relation (2) is the same as in SR theory.
So, a very important point to notice is that 
the speed of light $c$ is an invariant quantity in the MS model.
Another interesting
fact is that the  models described in \cite{cam}
have no finite upper bound on the energy of the photons though
they have a momentum upper bound.
On the other hand,
in the MS model\cite{mag1,pal}, though the dispersion relation for the
photons is unchanged, there is a finite upper bound on the
photon energy which is the Planck energy.
But
the problem of the
addition of momenta is not well established in DSR, so a classical
addition law is compatible with the model, but it is not the unique
possibility (see for example\cite{referee}).
\section{Density of states and Partition Function}
To study the thermodynamic behavior of photon gas, we
first find out  the partition function, as it enables us to calculate
the thermodynamics.
From the modified dispersion relation we find out the energy expression for the massless particle as,
\begin{equation}
 E=pc,
\end{equation}
where, $c$ is the velocity of photons
Considering a $d$-dimensional
  box of volume $V $ containing photon gas, we follow the
standard procedure as given in \cite{pathria}.
The number of microstates available to the system ($\sum$)
in the position range from $r$ to $r + dr$ and in the momentum
range from $p$ to $p + dp$ is given by
\begin{eqnarray}
 \sum=\frac{1}{h^d}\int\int d^d p d^d r,
 \end{eqnarray}
here, $h$ is the phase space volume of a single lattice and $\int\int d^d p d^d r$ is total volume of the phase space.
It should mentioned that, in SR theory the invariant quantities under Lorentz transformation are $\frac{d^d p}{E}$
and $Ed^dx$. As a result Eq. (21) remains invariant under Lorentz transformation.
In the case of $\kappa$-Lorentz transformation, we find out that, the NC phase space volume 
transforms as,
\begin{eqnarray}
 d^dx'd^dp'=\alpha ^d \gamma d^dx \frac{\gamma}{\alpha ^d}(1-\frac{\nu P^1}{E})d^dp= d^dxd^dp. \nonumber \\
\end{eqnarray}
So, following the way in \cite{pathria} we find out the density of states as,
\begin{eqnarray}
 g(E)dE=B(V,d) E^{d-1}dE,
\end{eqnarray}
where, 
\begin{eqnarray}
 B = \frac{2^{1-d} d \pi ^{-d/2} V}{\Gamma \left(\frac{d}{2}+1\right)h^d c^d}.
\end{eqnarray}
Here $\Gamma(j)=\int_0 ^\infty  x^{j-1}e^{-x} dx$. Putting $d=3$, one can find out that eq. (23) coincides with . \cite{pathria}.
Now in the grand canonical ensemble (GCE) the partition function for massless Bose gas can be written as\cite{pathria},
\begin{eqnarray}
log Z= -\sum_E log (1-e^{-\beta E}) .
\end{eqnarray}
Here, $\beta=\frac{1}{k_BT}$, $k_B$ is the Boltzmann constant,
and $T$ is the
temperature of the particle.
Changing the sum by integral we find out that, in SR theory\cite{pathria}
\begin{eqnarray}
 log Z= -\int_0 ^\infty  g(E) log (1-e^{-\beta E}) dE.
\end{eqnarray}
Here we have used Bose-Einstein distribution, which is the correct statistics for Bosons\cite{pathria} such as photon\cite{cam},
but we need to make a modification in eq. (26) to calculate thermodynamics of photons
in DSR theory using MS model.
Due to the presence of an energy upper
bound of particles $\kappa$ in the DSR theory
we have to make a modification in the above expression as below following the spirit of ref\cite{photon},
\begin{eqnarray}
 log Z= -\int_0 ^\kappa  g(E) log (1-e^{-\beta E}) dE.
\end{eqnarray}
Point to note that
the upper limit of integration is $\infty$ in eq (26), as
there is no upper bound of energy in the SR theory
but
the upper limit of
integration is $\kappa$ in (27).
 in the MS model which we
are considering, the photon dispersion relation is not
modified at all as $m=0$. But still there is modification in the partition 
function  due to the existence of an energy upper
bound of particles $\kappa$ in the theory.
In the limit $\kappa\longrightarrow \infty$, we get back the 
normal SR theory
results.
\section{Thermodynamics of photon gas}
So we have obtained the expression for partition function. In this section we calculate the  thermodynamics
 of photon gas in a Lorentz symmetry conserving DSR scenario.
\subsection{Free Energy}
In GCE, the free energy can be evaluated from the partition function,
\begin{eqnarray}
 F&=&-k_B T log Z \nonumber\\
 &=& k_B T\int_0 ^\kappa \rho(E) log (1-e^{-\beta E}) dE\nonumber\\
 &=&-\frac{2^{1-d}c^{-d} V \hbar^{-d} \beta^{-(d+1)} \zeta(1+d) \pi^{-\frac{d}{2}}}{\Gamma(\frac{d}{2}+1)} f(\kappa,d),\nonumber \\
\end{eqnarray}
here,  $f(\kappa,d)= \Gamma(1+d)-\Gamma(1+d,\kappa)$.
$\Gamma(j,k)$ is the incomplete gamma function, $\Gamma(j,k)=\int_k ^\infty l^{j-1} e^{-l}dl$.
We have removed the logarithm through an integration by parts. Most importantly the contribution of observer independent fundamental
energy scale enters through incomplete gamma function.\\ \\
Taking $\kappa \longrightarrow \infty$, we can find the SR result in $d$ dimension,
\begin{eqnarray}
 F=-\frac{2^{1-d}c^{-d} V \hbar^{-d} \beta^{-(d+1)} \zeta(1+d)\pi^{-d/2}}{\Gamma(\frac{d}{2}+1)} 
 \Gamma(d+1), \nonumber \\
\end{eqnarray}
putting, $d=3$ we can recover the familiar result for free energy of photon gas\cite{pathria},
\begin{equation}
 F=-\frac{V\pi^2}{45 \hbar^3 c^3}(k_BT)^4.
\end{equation}
We should point out that, the free energy as well as other thermodynamic quantities obtained Das and Roychowdhury 
\cite{photon} is unable to reproduce the known results\cite{pathria} of photon gas
in three dimensional space. Free energy of photons in three dimension has temperature dependency as
$F\propto - T^4$, but they obtained $F=N k_B T$, which is not correct but rather the result of classical ideal gas.
But this is not surprising as they have used Maxwell-Boltzmann distribution which is valid for
classical particles only.
\subsection{Internal energy}
Another important thermodynamic quantity internal energy $E$,
\begin{eqnarray}
 U&=&-\frac{\partial}{\partial \beta}log Z\nonumber\\
 &=&\frac{2^{1-d}c^{-d} V \hbar^{-d} d\beta^{-(d+1)} \zeta(1+d) \pi^{-\frac{d}{2}}}{\Gamma(\frac{d}{2}+1)} f(\kappa,d).\nonumber\\
\end{eqnarray}
Again as $\kappa$ tends to infinity we retrieve the SR results,
\begin{eqnarray}
 U=-\frac{2^{1-d}c^{-d} V \hbar^{-d}  \beta^{-(d+1)} d\zeta(1+d)\pi^{-d/2}}{\Gamma(\frac{d}{2}+1)}  \Gamma(d+1). \nonumber \\
\end{eqnarray}
In the case of $d=3$ the above equation coincides with known result as well\cite{pathria}.
\begin{figure}[h!]
\centering
\includegraphics[scale=0.9]{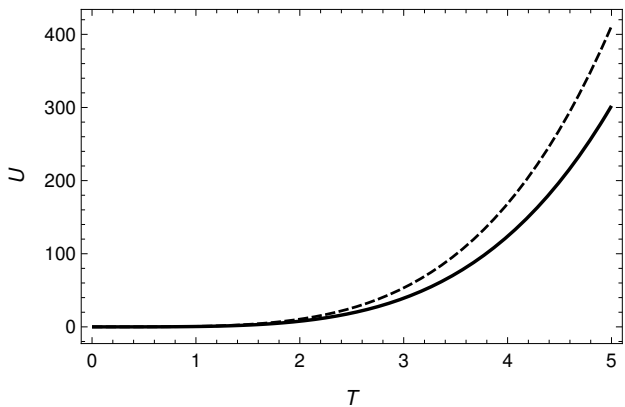}  
\caption{ Plot of internal energy of photon $E$ against
temperature T for both in the SR theory and in MS model in three dimensional space; the
dashed line corresponds to the SR theory result and the thick line
represents the corresponding quantity in our result. We have used
the Planck units and the corresponding parameters take the
following values  $\kappa= 5$ , $k_B=1$, $V = 1$,
$h = 1$ in this plot as well as in all other plots in the paper. In this
scale, $T =5$ is the Planck temperature. 
The dashed line corresponds to the SR theory result and the thick
line represents the quantity in the MS model.}
  \label{fig:boat1}
\end{figure}
In Fig. (1), we plotted internal energy of photon gas
against its temperature for both the case of MS model
and of SR theory in $d=3$.
It is clearly noticed from the plot that
the internal energy grows at a much slower rate in the case of our
result than in the SR theory and as temperature increases.
This is due to the fact
that Lorentz symmetry is further restricted in 
MS model. As a result of this, we expect to have a fewer
number of microstates and less internal energy in the MS model.
Note that, in both cases of SR and MS model internal energy has $T^4$
dependency but in MS model, the internal energy is less due to the presence of $\kappa$
through incomplete gamma function. 
It should also be mentioned that, internal energy in MS model is
related to free energy by-
\begin{eqnarray}
 F=-\frac{1}{d}U,
\end{eqnarray}
just as in SR theory\cite{pathria}. But this relation is not maintained in Lorentz violating model of Camacho and Marcias\cite{cam}.

\subsection{Entropy}
We can easily calculate the entropy from free energy,
\begin{eqnarray}
 S&=&-(\frac{\partial F}{\partial T})_{_T}\nonumber\\&=&
  \frac{2^{1-d}c^{-d} V \hbar^{-d} (d+1)\beta^{-(d+1)}T^{-1} \zeta(1+d) \pi^{-\frac{d}{2}}}{\Gamma(\frac{d}{2}+1)}f(\kappa,d).\nonumber\\ 
\end{eqnarray}
Again in the limit, $\kappa\longrightarrow\infty$ we find the $d$-dimensional result for SR theory,
\begin{eqnarray}
 S=\frac{2^{1-d}c^{-d} V \hbar^{-d} (d+1)\beta^{-(d+1)}T^{-1} \zeta(1+d) \pi^{-\frac{d}{2}}}{\Gamma(\frac{d}{2}+1)}\Gamma(d+1).  \nonumber\\
\end{eqnarray}
and coincides with known result when $d=3$ is chosen\cite{pathria}.
\begin{figure}[h!]
\centering
\includegraphics[scale=0.9]{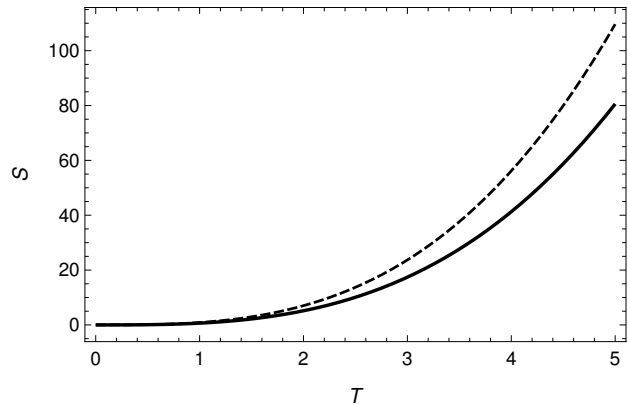}  
\caption{Plot of entropy of photon $S$ against
temperature T for both in the SR theory and in MS model with the same scaling as figure 1. 
The dashed line corresponds to the SR theory result and the thick
line represents the quantity in the MS model.} 
  \label{fig:boat1}
\end{figure}
In Fig. 2,  entropy against temperature
for  the MS model and
normal SR theory are plotted in three dimensional space.
As before, 
the entropy grows at a much slower rate in the case of our
result than in the SR theory and as temperature increases,
the entropy in our considered model deviates more from
the entropy in the SR theory.
This is well known that\cite{ca} the total number of microstates
available to a system is a direct measure of the entropy for
that system. Therefore our result merely reflects the fact
that due to the existence of an energy upper bound $\kappa$, the
number of microstates gradually decreases near Planck temperature.
But it should be noted that, this is not the case 
in the models with different dispersion relation where Lorentz symmetry is broken\cite{cam}.
In  \cite{cam} it is shown,
the entropy becomes larger as an unavoidable consequence
of this kind of Lorentz violation. But this is not the case in MS model, as Lorentz symmetry is well preserved here.
But nevertheless, in both types of model as $T\longrightarrow 0$ we find $S\longrightarrow0$,
indicating
Nernst postulate is always maintained, whether
if the Lorentz symmetry is broken or not.

\subsection{Pressure}
The pressure of photon gas,
\begin{eqnarray}
 P=(\frac{\partial F}{\partial T})_{_V}=
 \frac{2^{1-d}c^{-d}  \hbar^{-d} \beta^{-(d+1)} \zeta(1+d) \pi^{-\frac{d}{2}}}{\Gamma(\frac{d}{2}+1)} f(\kappa,d). \nonumber\\
\end{eqnarray}
In the limit $\kappa\rightarrow \infty$, we find out the pressure in $d$-dimensional SR theory is,
\begin{eqnarray}
P= \frac{2^{1-d}c^{-d}  \hbar^{-d} \beta^{-(d+1)} \zeta(1+d) \pi^{-\frac{d}{2}}}{\Gamma(\frac{d}{2}+1)}\Gamma(d+1). \nonumber \\
\end{eqnarray}
In $d=3$, we redeem the known result $P=\frac{\pi^2 k_B ^4}{45 \hbar^3c^3}T^4$.
The pressure of photon gas has a remarkable contribution in early universe cosmology,
as it was well dominated by photons\cite{L,L1}.
A very well known relation in SR theory between pressure and internal energy in three dimension is $P=\frac{U}{3V}$.
Comparing Eq. (32) and (36) we find out that same relation is also maintained in MS model.
In MS model, the $d$-dimensional relation between pressure and internal energy is,
\begin{eqnarray}
U=\frac{P}{d}V.  
\end{eqnarray}

But, this relation is not maintained in the other modified dispersion relation\cite{cam}
due to breakdown of Lorentz symmetry in their model. As it turns out, breakdown of Lorentz symmetry manifests as a repulsive interaction
Indeed, the
presence of a repulsive interaction (among the particles of a gas) entails the increase of the pressure, compared against
the corresponding value for an ideal gas. But we notice
the opposite in MS model 
in figure 1\footnote{As energy density $\rho$ is related to pressure $P$, by $\rho=\frac{P}{d}$, we did not make another plot for pressure.},
i.e. pressure increases in a slower rate with increasing temperature.

\subsection{Specific heat}
Specific
heat ($C_V$) is
defined as,
\begin{eqnarray}
 C_V&=&(\frac{\partial U}{\partial T})_{_V}\nonumber\\
 &=&  \frac{2^{1-d}c^{-d} V \hbar^{-d} d(d+1)T^{-1} \zeta(1+d) \pi^{-\frac{d}{2}}}{\beta^{d+1}\Gamma(\frac{d}{2}+1)}f(\kappa,d). \nonumber \\
\end{eqnarray}
Also, when $\kappa\rightarrow \infty$,
we recover the $d$-dimensional result for specific heat of photons.
\begin{eqnarray}
 C_V&=&(\frac{\partial U}{\partial T})_{_V}\nonumber \\
  &=& \frac{2^{1-d}c^{-d} V \hbar^{-d} d(d+1)T^{-1} \zeta(1+d) \pi^{-\frac{d}{2}}}{\beta^{d+1}\Gamma(\frac{d}{2}+1)} \Gamma(d+1).   \nonumber \\
\end{eqnarray}
The above equation coincides with known result when $d=3$.
In our calculation, $C_V$ in both MS model and 
SR theory has $T^d$ dependency in $d$-dimension, unlike \cite{photon} who have 
reported constant value of specific heat in SR theory.
This is completely wrong as it is well established that, $C_V$ of photon gas
has $T^3$ dependency in three dimension\cite{pathria}. The constant specific heat is rather a result of
ideal non relativistic classical gas. In our study we have noticed from eq. (34) and (39) that,
specific heat is related to entropy as,
\begin{equation}
S=\frac{C_V}{d} . 
\end{equation}
This relation is a well
established result is SR\cite{pathria}.
So above relation along with figure 2 dictates that like the other thermodynamic quantities in MS model,
specific heat changes in a slow  rate with temperature compared to SR theory.
On the other hand, we find the opposite in Lorentz violating model of Camacho and Macias\cite{cam}.
Their interpretation of 
 the breakdown of Lorentz symmetry as
the appearance of a repulsive interaction,
results a larger specific heat in SR theory.
As $C_V$ is a measurable quantity\cite{cam}, which in principle, could be employed in the experimental
quest for violations of Lorentz symmetry, our present calculation is a significant  justification 
of the theoretical status.
\section{Conclusion}
In this paper we successfully calculated the
thermodynamics
 of photon gas in a theory where an observer-
independent fundamental energy scale is present.
The most important  part of the present work is the  derivation
of  partition function in the MS model in arbitrary dimensions. Due to
 the deformed dispersion relation, this
task becomes highly nontrivial to find the partition function analytically. However, for photons, we
find out an exact analytic expression for the partition function, enabling us to calculate thermodynamic
quantities such as the free energy, pressure, entropy, internal
energy, and specific heat for the MS model, and compare
them with the known results of SR theory in three dimensional space.
It should be noted that the  influence of the Planck scale enters through incomplete gamma function.
 As expected, our results match with the known results\cite{pathria} of SR theory in the limit 
  $\kappa\rightarrow\infty$ unlike Ref\cite{photon}.
But due to the presence of an invariant energy upper
bound in this theory, the microstates can avail energies only up
to a finite cutoff, whereas in the SR theory, the microstates can
attain energies up to infinity.
As a result,
the
number of the microstates in this MS model is less than that
in SR theory.
This is clear from  the result we obtained for entropy (figure 2) as entropy indicates the total number of the microstates available.
This
 happens since Lorentz symmetry is not broken  but is rather more restricted
 in the MS
model. Just the opposite happens in the model\cite{cam}, where
Lorentz symmetry is not preserved. It is
shown in \cite{cam} that the number of the microstates available to the corresponding equilibrium
state grows, compared to the SR theory.
The entropy becomes larger as an unavoidable consequence of
this kind of Lorentz violation.
Additionally,  the breakdown of Lorentz symmetry entails a larger
value of pressure, internal energy or any other thermodynamic quantity\cite{cam} compared to the SR results.
As noticed, an entirely different scenario is obtained in the current study with MS model.
But it is very intriguing to note that Nernst postulate, i.e. the third law of thermodynamics is 
maintained in both the MS model and Lorentz violating study of Camacho and Marcias.
So in conclusion, in the MS model, 
 where the Lorentz algebra is still
intact 
in the presence of observer independent fundamental energy scale,
yields that the thermodynamic quantities grow slowly against temperature compared to the SR theory whereas in the Lorentz violating study, they tend to increase more quickly with temperature than in the SR theory. Also
some very well established relations\cite{pathria} among different thermodynamic quantities of photons in SR theories are eq. (33), (38) and (41). These equations are valid in the MS model but not in  \cite{cam}.
These are the key differences in the study of photon thermodynamics in Lorentz symmetry violating and Lorentz symmetry obeying models, which will play an important role in examining space-time structure near Planck scale\cite{C23}. It would be interesting if these key differences are also maintained in
the case of massive quantum gases.

Since the modification of the dispersion relation has changed the thermodynamics of photon drastically, we need to explicitly examine the thermodynamics of massive quantum gases in the MS model.
Since
the so-called Bose-Einstein condensation and Fermi degeneracy are
purely bosonic and fermionic effects respectively, then we
may wonder what happens to this feature if we introduce the generalization to
(2) for massive particles. It would certainly change the condensation temperature for Bose gas as well as Fermi temperature for Fermi gas. The former case is intriguing  in scalar field dark matter model, where dark matter particle is a spin-0 boson\cite{L,L1}. But the later case is important  since the Chandrasekhar mass-radius relation\cite{C} for white
dwarfs is a direct consequence of the fermionic statistics.
Hence we
expect a modification in these studies
due to the presence of   an  observer  independent  fundamental energy scale. 
Besides, one can study the cosmological
aspects of the MS model using the Friedmann equations.
But, this still remains another open issue to be further studied.
\begin{acknowledgments}
 The work is supported by grants from Commonwealth Agency, UK (BDCS-2015-20). 
MMF would like to thank Mr. Alamgir Al Faruqi for his cordial hospitality during MMF's stay  in London, UK.
Thanks to Fathema Farjana, David Enrique Rodriguez Bernal, Constantin Stan Weisser and Liana Islam    for their effort to help us
present the manuscript.
\end{acknowledgments}


\begin{thebibliography}{5}
\bibitem{mag1} J. Magueijo and L. Smolin, Phys. Rev. Lett. 88, 190403
(2002);
\bibitem{mag2} J. Magueijo and L. Smolin, Phys. Rev. D 67, 044017 (2003).
\bibitem{Am} G. Amelino-Camelia, Nature (London) 418, 34 (2002);
\bibitem{Am1} G. Amelino-Camelia, Phys. Lett. B 510, 255 (2001).
\bibitem{Am2}C. Rovelli and L. Smolin, Nucl. Phys. B442, 593 (1995);
B456, 734(E) (1995); G. Amelino-Camelia, Mod. Phys.
Lett. A 17, 899 (2002).
\bibitem{g}J Polchinski, hep-th/9611050; S. Carlip, Rep. Prog. Phys. 64, 885 (2001).
\bibitem{k} J. Kowalski-Glikman and S. Nowak, Phys. Lett. B
539, 126 (2002); Classical Quantum Gravity 20, 4799
(2003).
\bibitem{pal}S. Ghosh and P. Pal, Phys. Rev. D 75, 105021 (2007).
\bibitem{m}J. Magueijo, Phys. Rev. D 63, 043502 (2001).
\bibitem{emt} S. Das, S. Ghosh, and D. Roychowdhury, Phys. Rev. D 80,
125036 (2009).
\bibitem{malu} Nitin Chandra and Sandeep Chatterjee, Phys. Rev. D 85, 045012 (2012).
\bibitem{malu2}Xinyu Zhang, Lijing Shao, Bo-Qiang Ma, Astroparticle Physics 34 (2011) 840-845
\bibitem{pp}S. Doplicher, K. Fredenhagen, and J. E. Roberts, Phys.
Lett. B331, 39 (1994).
\bibitem{curved} Ahmed Farag Ali, Mir Faizal, Mohammed M. Khalil,  Nucl.Phys. B894 (2015) 341,  Seyed Hossein Hendi, Mir Faizal, Phys. Rev. D 92, 044027 (2015),   Ahmed Farag Ali, Mir Faizal, Barun Majumder, Ravi Mistry, Int.J.Geom.Meth.Mod.Phys. 12 (2015), 1550085, Ahmed Farag Ali, Mir Faizal, Mohammed M. Khalil, JHEP 1412 (2014) 159,
Ahmed Farag Ali, Mir Faizal, Mohammed M. Khalil, Phys.Lett. B743 (2015) 295,  Amani Ashour, Mir Faizal, Ahmed Farag Ali, Fayçal Hammad,   Eur. Phys. J. C76 (2016)  264, Ahmed Farag Ali, Mir Faizal, Barun Majumder,  Europhys.Lett. 109 (2015) 20001
\bibitem{J. Kowalski-Glikman}J. Kowalski-Glikman,
  Lecture Notes in Physics pp 131-159, 
  J. Kowalski-Glikman,
Phys. Lett. A 299 (2002) 454
\bibitem{kkk}Sudipta Das, Souvik Pramanik and Subir Ghosh,  SIGMA 10 (2014), 104
\bibitem{lll} C Barceló, S Liberati, M Visser, Living Rev. Relativity, 14, (2011), 3 
\bibitem{nc}  J. Kowalski-Glikman and S. Nowak, Phys. Lett. B
539, 126 (2002) 
\bibitem{tc}F. Briscese et al,  EPL 98 60001 (2012).
\bibitem{rainbow}João Magueijo and Lee Smolin, Class. Quantum Grav. 21 1725 (2004).
\bibitem{dragon}A. A. Deriglazov, Phys.Lett. B603 (2004) 124-129.
\bibitem{qg}M. Grether, M. de Llano, and George A. Baker, Jr.
Phys. Rev. Lett. 99, 200406
\bibitem{L}L. Arturo Urena-Lopez, JCAP01 (2009) 014.
\bibitem{L1}Saurya Das and Rajat K Bhaduri, Class. Quantum Grav. 32 (2015) 105003
\bibitem{mmf}Mir Mehedi Faruk and Md Muktadir Rahman, J. Stat. Mech. (2016) 033117.
\bibitem{photon}Sudipta Das and Dibakar Roychowdhury, Phys. Rev. D 81, 085039 (2010)
\bibitem{peskin} M. E. Peskin and D. V. Schroeder, An Introduction to Quantum Field Theory,
Addison-Wesley Publishing Company (1995).
\bibitem{pathria}R. K. Pathria, Statistical Mechanics (Butterworth-
Heinemann, Oxford, 1996)
\bibitem{ca} Callen, H.B.,Thermodynamics and an introduction to thermostatistics. Wiley,NewYork (1985)
\bibitem{l} N. R. Bruno, G. Amelino-Camelia, and J. Kowalski-
Glikman, Phys. Lett. B 522, 133 (2001).
\bibitem{cam}A. Camacho and A. Macius, Gen. Relativ. Gravit. 39, 1175
(2007).
\bibitem{de}O. Bertolami and C. A. D. Zarro, Phys. Rev. D 81, 025005
(2010); M. Gregg and S. A. Major, Int. J. Mod. Phys. D 18,
971 (2009).
\bibitem{exp} J. Ellis et al., Astrophys. J. 535, 139 (2000);
J. Ellis, N. E. Mavromatos, and D. Nanopoulos, Phys. Rev.
D 63, 124025 (2001)
\bibitem{exp1}R. Gambini and J. Pullin, Phys. Rev. D 59, 124021 (1999).
\bibitem{exp2}J. Alfaro et al., Phys. Rev. Lett. 84, 2318 (2000).
\bibitem{exp3}J. Alfaro, Phys. Rev. D 72, 024027 (2005); 
E. Waxman , J. Bahcall, Phys. Rev. Lett. 78, 2292 (1997);
\bibitem{velo}S. Hossenfelder, Classical Quantum Gravity 23, 1815
(2006).
\bibitem{C}P.H. Chavanis
Phys. Rev. D 76, 023004.
\bibitem{C23}J. Alfaro, Phys. Rev. D 72, 024027 (2005);
E. Waxman, J. Bahcall, Phys. Rev. Lett. 78, 2292 (1997);
\bibitem{referee}
Carmona et al. Phys. Rev. D86, 084032.
\end{thebibliography}

\end{document}